\documentstyle[12pt]{article}
\def\beq{\begin{equation}}
\def\eeq{\end{equation}}
\def\bea{\begin{eqnarray}}
\def\eea{\end{eqnarray}}
\def\nn{\nonumber}
\def\ba{\begin{array}}
\def\ea{\end{array}}
\def\d{\partial}
\def\v{\vert}
\def\l{\langle}
\def\r{\rangle}
\def\one{1\hskip -1mm{\rm l}}
\def\P{{\rm l}\hskip -.85mm{\rm P}}

\begin{document}
% \begin{flushright} 
% \end{flushright}
%\rightline{}
%\rightline{February 1998}
\baselineskip16pt
\smallskip
\begin{center}
{\large \bf \sf
Multi-parameter deformed and nonstandard $Y(gl_M)$ Yangian symmetry \\
in integrable variants of Haldane-Shastry  spin chain }\\

\vspace{1.75 cm}

{\sf B. Basu-Mallick\footnote{ 
E-mail address: biru@monet.phys.s.u-tokyo.ac.jp } }

\bigskip

{\em Department of Physics, Faculty of Science, University of Tokyo, \\
 Hongo 7-3-1, Bunkyo-ku, Tokyo 113, Japan }

\end{center}

\vspace {2.5 cm}
\baselineskip=20pt
\noindent {\bf Abstract }

By using  `anyon like' representations of permutation algebra, 
which pick up nontrivial phase factors while interchanging the spins of two
lattice sites, we construct some integrable variants of Haldane-Shastry (HS)
spin chain. Lax equations for these spin chains allow us to find
out the related conserved quantities. However, it turns out that such spin 
chains also possess a few additional conserved quantities which are 
apparently not derivable from the Lax equations. Identifying these additional 
conserved quantities, and the usual ones related to Lax equations, 
with different modes of a  monodromy matrix,  it is shown that the 
above mentioned HS like spin chains exhibit multi-parameter deformed and
`nonstandard' variants of $Y(gl_M)$ Yangian symmetry.

\newpage

\baselineskip=22pt
\noindent \section {Introduction }
\renewcommand{\theequation}{1.{\arabic{equation}}}
\setcounter{equation}{0}

\medskip
One dimensional spin chains  as well as
dynamical systems with long ranged interactions
and their close connection with diverse subjects like Yangian algebra,
random matrix theory, fractional statistics, quantum Hall effect etc. have
attracted a lot of attention in recent years [1-15].
In particular it is found that,  commutation relations among the
conserved quantities of spin Calogero-Sutherland (CS)
model, given by the Hamiltonian
\beq
H_{CS} ~=~  -{1\over 2} \sum_{i=1}^N ~ \left ( { \d \over \d x_i }
\right )^2 ~+
{ \pi^2 \over L^2 }~ \sum_{i<j}~{   \beta ( \beta + P_{ij}  )  \over
\sin^2 { \pi \over L} (x_i -x_j)  }~,
\label {a1}
\eeq
where $\beta $ is a coupling constant and
$P_{ij}$ is the permutation operator interchanging the `spin' components 
(taking $M$ possible values) of $i$-th and $j$-th particles, 
generate the  $Y(gl_M)$
Yangian algebra [5]. Furthermore, commutation relations
among the conserved quantities of Haldane-Shastry (HS) spin chain,
 which may be obtained by taking the `static limit' 
of spin CS model, also lead to this $Y(gl_M)$
Yangian algebra [2]. 

However, it is recently found that the permutation algebra
given by
\beq
  {\cal P}_{ij}^2 ~=~ 1,
~~~{\cal P}_{ij}{\cal P}_{jl} ~=~ {\cal P}_{il}{\cal P}_{ij} ~=~
{\cal P}_{jl} {\cal P}_{il} ~, ~~~[ {\cal P}_{ij},
{\cal P}_{lm} ] ~=~ 0~ ,
\label {a2}
\eeq
($ i, ~j,~l,~m $  being all different indices), admits a novel 
class of  `anyon like' representations on the internal 
space of spin CS model [16].
Such an `anyon like' representation (${\tilde P}_{ij}$) 
acts on the internal space of all particles as
\beq
{\tilde P}_{ij} \,
 \v \alpha_1 \alpha_2 \cdots \alpha_i \cdots \alpha_j
\cdots \alpha_N \r ~=~ e^{ i \Phi (\alpha_i , \alpha_{i+1} , \cdots ,
\alpha_j ) } \, \v \alpha_1
\alpha_2 \cdots \alpha_j \cdots \alpha_i \cdots \alpha_N \r \,   , ~
\label {a3} \eeq
where  $\alpha_i~ (\in [1,2,\cdots ,M] )$ denotes a spin variable
 and $ \Phi (\alpha_i , \alpha_{i+1} , \cdots , \alpha_j ) $
is a real function of $(j-i+1)$ number of such spin variables. So these
`anyon like' representations pick up nontrivial phase factors
while interchanging the spin of two particles and reduce to
the standard representation of permutation algebra ($P_{ij}$)
at the limiting case $\Phi = 0$. Many representations of permutation algebra,
including the type  (\ref {a3}),
can be constructed systematically by using the relation given by [17]
\beq
{\tilde P}_{i,i+1}  =  Q_{i,i+1}P_{i,i+1}  , ~
{\tilde P}_{ij}  =  S_{ij} {\tilde P}_{j-1,j}S_{ij}^{-1} = 
\left( Q_{i,i+1}  Q_{i,i+2}  \cdots  Q_{ij} \right)   P_{ij} 
\left(  Q_{j-1,i} Q_{j-2,i}  \cdots   Q_{i+1,i}  \right) \, ,
\label {a4}
\eeq
where $j >i+1$,  $ S_{ij} = {\tilde P}_{i,i+1} {\tilde P}_{i+1,i+2} \cdots  
 {\tilde P}_{j-2,j-1} $ and  $Q_{ik}$, which acts like a matrix $Q$ on the 
direct product of $i$-th and $k$-th spin spaces (but acts trivially on 
all other spin spaces), is obtained by solving the following
two equations:
\beq
Q_{ik} \, Q_{il} \, Q_{kl} ~=~ Q_{kl} \, Q_{il} \, Q_{ik}
\, , ~~~~~ Q_{ik} \, Q_{ki}  ~=~ \one \, .
\label {a5}
\eeq
It is easy to check that the 
above equations, satisfied by the $Q$ matrix, are in fact 
sufficient conditions for ${\tilde P}_{ij}$ (\ref {a4}) being a  
a valid representation of permutation algebra (\ref {a2}). Consequently,
by inserting any solution of eqn.(\ref {a5}) to (\ref {a4}), we can 
generate a representation of permutation algebra.  For 
the case $Q= \one $, which is the simplest solution 
of eqn.(\ref {a5}), ${\tilde P}_{ij}$ (\ref {a4}) reproduces the standard 
representation $P_{ij}$. Moreover,   concrete examples of 
`anyon like' representations (\ref {a3}) can also be obtained 
in a similar way from a class of nontrivial solutions of 
eqn.(\ref {a5}) [17].
 
Interestingly,  one may construct integrable variants of 
spin CS Hamiltonian  (\ref {a1}) as
\beq
{\cal H}_{CS} ~=~  -{1\over 2} \sum_{i=1}^N ~ \left ( { \d \over \d x_i }
\right )^2 ~+ { \pi^2 \over L^2 }~ \sum_{i<j}~{   
\beta ( \beta + {\tilde P}_{ij}  )  \over
\sin^2 { \pi \over L} (x_i -x_j)  }~,
\label {a6}
\eeq
${\tilde P}_{ij} $ being any possible `anyon like' representation of 
permutation algebra, and show that such Hamiltonians 
can be solved exactly  by
taking appropriate projections of the eigenfunctions of Dunkl operators
[16]. However, it turns out that the spectra of
spin CS Hamiltonians (\ref {a6}) might differ 
considerably from that of their standard counterpart (\ref {a1}).
Furthermore it is found that, the symmetry of a spin CS model like 
(\ref {a6}) crucially depends on
 the choice of corresponding representation ${\tilde P}_{ij}$, 
and may be given by a multi-parameter deformed or `nonstandard' 
variant of $Y(gl_M)$ Yangian algebra [17].

It is evident that,
in analogy with the case of spin CS models (\ref {a6}),
one may also define novel variants of HS spin chain as
\beq
{\cal H}_{HS} ~=~  \sum_{ 1 \leq i <j \leq N } \,
  { 1 \over 2 \sin^2 {\pi \over N}(i-j)} \left ( {\tilde P}_{ij} - 1 \right ) 
\, ,
\label {a7}
\eeq
where ${\tilde P}_{ij}$ is any possible `anyon like' representation 
of permutation algebra which can be obtained through the relation
(\ref {a4}).  For the limiting case
${\tilde P}_{ij}= P_{ij}$, ${\cal H}_{HS}$  (\ref {a7})
reproduces the standard HS spin chain [1,6] which exhibits the $Y(gl_M)$
 Yangian symmetry.  So, it is natural to ask whether 
the spin model (\ref {a7}) would also be quantum integrable 
and respect some Yangian like symmetry for all possible choice of 
corresponding ${\tilde P}_{ij}$.  In the present article, our aim is to
answer this question by exploiting an
intimate connection between spin CS model and HS spin chain. So in sec.2, 
we start with the known Lax operators of spin CS Hamiltonian (\ref {a6})
and show that the `static limit' of such operators would easily
lead to the Lax pair of HS like spin chain (\ref {a7}). 
Lax equations for the spin chain (\ref {a7}) immediately
allow us to write down the associated conserved quantities. Subsequently,
in sec.3, we find that the spin chain (\ref {a7}) would also
possess a few additional conserved quantities which are not 
derivable from the Lax equations. Identifying these additional 
conserved quantities, and the usual ones related to Lax equations, 
with different modes of a  monodromy matrix,  we are able to demonstrate
that the spin chain (\ref {a7}) would respect a multi-parameter 
deformed or `nonstandard' variant of $Y(gl_M)$ Yangian symmetry. It may be
noted that, while deriving the above mentioned result, we do not 
assume any specific form of the `anyon like' representation 
${\tilde P}_{ij}$ and develop a rather general method for 
 finding out the symmetry of spin chain (\ref {a7}). However, 
at the end of sec.3, we finally use a particular class of 
${\tilde P}_{ij}$ for generating some concrete examples of spin chains 
(\ref {a7}) and apply our general method for 
 finding out the explicit forms of related 
symmetry algebras.  Sec.4 is the concluding section.
\vspace{1cm}

\noindent \section { Lax pairs and conserved quantities for
HS like spin chains }
\renewcommand{\theequation}{2.{\arabic{equation}}}
\setcounter{equation}{0}

\medskip
In this section we like to start with the known Lax operators 
of spin CS model (\ref {a6})
and explore how they might be used to generate the Lax pair
of HS like spin chain (\ref {a7}).  
For this purpose, it is convenient to rewrite 
${\cal H}_{CS}$ (\ref {a6}) as a power series expansion of the coupling
constant $\beta$:
\beq
{\cal H}_{CS}~=~ {2\pi^2 \over L^2} \, \left ( \, {\cal H}_0 + 
\beta {\cal H}_1 + \beta^2 {\cal H}_2 \, \right ) \, ,
\label {b1}
\eeq
where ${\cal H}_0 = \sum_{j=1}^N (z_j { \d \over \d z_j  })^2 \, ,$ 
${\cal H}_1 = 2 \sum_{i<j} \theta_{ij} \theta_{ji} {\tilde P}_{ij} \, ,$ 
${\cal H}_2 = 2 \sum_{i<j} \theta_{ij} \theta_{ji} \,$ and   $z_j,~
\theta_{ij}$ are defined as:
$ z_j = e^{{ 2\pi i \over L} x_j} $, $ \theta_{ij} =
{ z_i \over z_i - z_j } $. Lax operators 
associated with this spin CS Hamiltonian may be given by [17]
\beq 
{\cal L}~=~ {\cal L}_0 + \beta {\cal L}_1 \, , ~~
{\cal M}~=~ {2\pi^2 \over L^2} \beta {\cal M}_1 \, , ~~
\label {b2}
\eeq
where  ${\cal L}_0  $, ${\cal L}_1 $ and 
${\cal M}_1 $ are $N\times N$ matrices with operator valued elements:
$$
\left ( {\cal L}_0 \right )_{ij} = \delta_{ij} z_j { \d \over \d z_j },~
\left ( {\cal L}_1 \right )_{ij} = 
\left ( 1 - \delta_{ij} \right ) \theta_{ij} {\tilde P}_{ij},~ 
\left ( {\cal M}_1 \right )_{ij} = 
-  2 \delta_{ij} \sum_{k \neq i} h_{ik} {\tilde P}_{ik}
+ 2 \left ( 1 - \delta_{ij} \right ) h_{ij} {\tilde P}_{ij}, 
$$
  $h_{ij} = \theta_{ij} \theta_{ji} $ 
and it is assumed that $ {\tilde P}_{ij}$ is same as ${\tilde P}_{ji} $.
It may be noted  that, the Lax pair (\ref {b2}) 
can be obtained from the Lax pair 
of usual spin CS model (\ref {a1}) [5,7] by simply substituting 
${\tilde P}_{ij}$ to the place of $P_{ij}$. However, unlike the 
usual case, it is necessary to define some other nontrivial operators 
for obtaining all Lax equations related to the  spin CS model (\ref {a6}). 
For this purpose, we consider the $Q$ matrix which is a solution of
eqn.(\ref {a5}) and generates 
${\tilde P}_{ij}$ through the relation (\ref {a4}). By using this $Q$ 
matrix, one may define a set of operators as
\beq
 {\tilde X}_i^{ \alpha \beta } ~=~ Q_{i,i+1} Q_{i,i+2} \cdots
 Q_{iN}  \, X_i^{\alpha \beta } \,
  Q_{i1} Q_{i2} \cdots Q_{i,i-1} \, ,
\label {b3}
\eeq
where $X_i^{\alpha \beta }$ acts like $ \v \alpha \r \l \beta \v $
on the spin space associated with $i$-th particle and 
leaves the spin spaces associated with all other particles 
untouched. It is clear that, for the limiting 
case $Q= \one $ (i.e., when ${\tilde P}_{ij}=P_{ij}$),
 ${\tilde X}_i^{\alpha \beta }$ (\ref {b3}) reduces to the local operator 
 $ X_i^{\alpha \beta }$. However, the above defined 
 ${\tilde X}_i^{\alpha \beta }$s would generally represent a set of
highly nonlocal spin dependent operators. Moreover, by applying
eqn.(\ref {a5}), it can be shown that 
 ${\tilde X}_i^{\alpha \beta }$  and ${\tilde P}_{ij}$ (\ref {a4}) obey the 
simple algebraic relations:
\beq
 {\tilde P}_{ij}{\tilde X}_i^{ \alpha \beta } ~=~ 
 {\tilde X}_j^{ \alpha \beta } {\tilde P}_{ij} \, .
\label {b4}
\eeq
By using the above equation, the permutation algebra (\ref {a2})
  and the canonical commutation relations: $ [{\d \over \d z_i }, z_j ]
= \delta_{ij}$,   it is easy to see that the Hamiltonian 
${\cal H}_{CS}$ (\ref {b1}), Lax pair ${\cal L}$ and ${\cal M}$ (\ref {b2}),
and the operators ${\tilde X}_i^{ \alpha \beta }$ (\ref {b3}) satisfy the
Lax equations given by 
\bea
&\left [ {\cal H}_{CS} , 
 {\cal L}_{ij} \right ] = \sum_{k=1}^N
 \left (  {\cal L}_{ik} {\cal M}_{kj} -
 {\cal M}_{ik} {\cal L}_{kj}   \right )  , ~ \left [{\cal H}_{CS} ,
  {\tilde X}_j^{\alpha \beta } \right ] = \sum_{k=1}^N
 {\tilde X}_k^{\alpha \beta } {\cal M}_{kj}   ,~ \sum_{k=1}^N
{\cal M}_{jk} = 0  \, , ~~ ~~~~~~~~~~~~~~~~~~~~~~~\nn \\
& \hskip 10.40 true cm
 \nn  (2.5a,b,c)
\eea
\addtocounter{equation}{1}
With the help of these Lax equations, one can easily derive the conserved 
quantities of spin CS model (\ref {a6}) and show that 
the commutation relations among such conserved 
quantities would lead to a multi-parameter dependent or `nonstandard' variant
of $Y(gl_M)$ Yangian algebra [17]. 

However, for our present purpose of finding out the 
Lax operators associated with HS like spin chain (\ref {a7}), 
it is useful to  observe that eqns.(2.5a,b,c) are  satisfied for 
any value of the coupling constant $\beta $ which appears in the 
Hamiltonian (\ref {b1}) as well as the Lax pair (\ref {b2}).
 Consequently, by inserting (\ref {b1}) and (\ref {b2}) to 
eqns.(2.5a), (2.5b) and (2.5c), and comparing the coefficients of 
$\beta^2$, $\beta$ and $\beta $ respectively
 from both sides of these equations,
 we can derive the following set of independent relations:
\bea
&\left[ \, {\cal H}_1 \, , \,  \left( {\cal L}_1 \right)_{ij} \, \right ] 
~+~ \left [ \, {\cal H}_2 ,  \, \left({\cal L}_0 \right)_{ij} \, \right] 
~=~ \sum_{k=1}^N \, \left \{ \, \left( {\cal L}_1 \right)_{ik} 
\left({\cal M}_1 \right)_{kj} \,  - \, \left({\cal M}_1 \right)_{ik} 
\left({\cal L}_1 \right)_{kj}   \right \} \, , \nn ~&~~(2.6a) \\ 
& \left [ {\cal H}_1 ,  {\tilde X}_j^{\alpha \beta } \right ] ~=~ \sum_{k=1}^N
 {\tilde X}_k^{\alpha \beta } ({\cal M}_1)_{kj}  \, ,
~ \sum_{k=1}^N \left({\cal M}_1 \right)_{jk} = 0  \, . \nn  ~~&(2.6b,c) 
\eea
\addtocounter{equation}{1}
It is curious to notice that, the form of eqn.(2.6a) would coincide 
with the form of Lax equation (2.5a), provided we assume the validity
 of following simple condition: 
\beq
  \left [ \, {\cal H}_2 \, ,  \, ({\cal L}_0)_{ij} \, \right ] ~=~ 0 \, .
\label {b7}
\eeq
However, by substituting the explicit form of 
  $ {\cal H}_2 $ and $({\cal L}_0)_{ij} $ to the above equation, we find
 that it reduces to
\beq
 \sum_{ k \neq j } \, \theta_{jk} \theta_{kj} \, \left ( \,  
  \theta_{jk} \,- \, \theta_{kj} \, \right ) \, = \, 0  \, . 
\label {b8}
\eeq
It is well known that eqn.(\ref {b8}) admits  the  solution:
$z_k = \omega^k$ with $ \omega = e^{2 i \pi \over N}$ [5] and 
leads to the `static limit', where the 
particles of spin CS model 
are frozen at equidistant positions on the arc of an unit circle. 
Thus it is clear that, by imposing the  `static limit',
 eqns.(2.6a,b,c) can be recast in the form of Lax 
equations where  ${\cal H}_1$ plays the role of Hamiltonian and 
 the operators $ {\cal L}_1, ~ {\cal M}_1$ play the role of 
 related Lax pair. It is interesting to notice further that,
at this static limit, the operator ${\cal H}_1$  would coincide
 with the Hamiltonian (\ref {a7}) up to an additive 
constant factor. Consequently,
the static limit of eqns.(2.6a,b,c) should give us
the desired Lax equations for HS like spin chain (\ref {a7}).
Therefore, we can explicitly write down the Lax equations associated
 with the Hamiltonian (\ref {a7}) as
\bea
&\left [ \, {\cal H}_{HS}  \, , 
\,   \left ({\cal L}_{HS} \right )_{ij} \, \right ] ~=~ \sum_{k=1}^N
 \, \left  \{ \, \left ({\cal L}_{HS} \right )_{ik} 
\left ({\cal M}_{HS} \right )_{kj} \, - \,
 \left ({\cal M}_{HS} \right )_{ik} \left ({\cal L}_{HS} \right )_{kj}
 \right \} \, , \nn \\ &\left [ \, {\cal H}_{HS} \,
,  \, {\tilde X}_j^{\alpha \beta } \, \right ] ~=~ \sum_{k=1}^N \,
 {\tilde X}_k^{\alpha \beta } \, 
\left ({\cal M}_{HS} \right )_{kj}  \, ,~ \sum_{k=1}^N
\left ({\cal M}_{HS} \right )_{jk} = 0  \, , ~\label {b9}
\eea
where  the elements of Lax pair are given by
\beq
\left ( {\cal L}_{HS} \right )_{ij} =
\left ( 1 - \delta_{ij} \right ) \theta'_{ij} {\tilde P}_{ij} \, ,~~
\left ( {\cal M}_{HS} \right )_{ij} = 
 -  2 \delta_{ij} \sum_{k \neq i} h'_{ik} {\tilde P}_{ik}
+ 2 \left ( 1 - \delta_{ij} \right ) h'_{ij} {\tilde P}_{ij} \, ,
\label {b10}
\eeq
with $ \theta'_{kj} = { \omega^k \over \omega^k - \omega^j }$ and 
 $h'_{kj} = \theta'_{kj} \theta'_{jk} $. Now,
by using the Lax equations (\ref {b9}), it is easy to demonstrate that 
the operators given by
\beq
 T_n^{ \alpha \beta } ~=~ \sum_{i,j=1}^N \,
 {\tilde X}_i^{\alpha \beta }
\left (  {\cal L}_{HS}^n \right )_{ij} \, ,
\label {b11}
\eeq
where $n \in [0,1, \cdots , \infty ]$ and $\alpha , \, \beta \, \in 
[ 1, \cdots , M] $, would commute with the Hamiltonian (\ref {a7}).
 
Thus, by imposing the `static limit' on the known Lax 
equations of spin CS model (\ref {a6}),  
we are able to derive here the Lax equations and conserved quantities 
associated with any spin chain which can be written in the 
form (\ref {a7}). It is evident 
that, at the limiting case $Q= \one $,
 (\ref {b11}) would reproduce the local conserved quantities of
 standard HS spin chain. However, for any nontrivial choice
of ${\tilde P}_{ij}$ or corresponding $Q$ matrix, 
eqn.(\ref {b11}) would  give us
 highly nonlocal type of conserved quantities associated with the
 spin chain (\ref {a7}).  In the next section, 
our aim is to find out the algebraic relations satisfied by
such conserved quantities and establish their connection with
the Yang-Baxter equation.

\vspace{1cm}

\noindent \section { Extended $Y(gl_M)$ Yangian symmetry in HS like
spin chains }
\renewcommand{\theequation}{3.{\arabic{equation}}}
\setcounter{equation}{0}

\medskip

We have already mentioned that,  commutation relations between 
the conserved quantities of HS spin chain yield the $Y(gl_M)$
Yangian algebra [2]. This  $Y(gl_M)$ Yangian algebra [18,19]
can be defined through the operator valued elements of an
$M\times M$ dimensional monodromy matrix $T^0(u)$,
which obeys the quantum Yang-Baxter equation (QYBE)
\beq
R_{00'}(u-v) \left ( T^0(u) \otimes \one \right )
 \left ( \one \otimes  T^{0'}(v) \right ) ~=~
 \left ( \one \otimes  T^{0'}(v) \right )
 \left ( T^0(u) \otimes \one \right )  R_{00'}(u-v) \, .
\label {c1}
\eeq
Here $u$ and $v$ are  spectral parameters and
the $M^2 \times M^2$ dimensional rational $R(u-v)$ matrix,  having 
$c$-number valued elements, is taken as
\beq
R_{00'} (u-v)  ~=~   (u-v) \, \one  \, + \, P_{00'} \, .
\label {c2}
\eeq
Associativity of $Y(gl_M)$
algebra is ensured from the fact that the above $R$ matrix 
satisfies Yang-Baxter equation (YBE) given by
\beq
  R_{00'} (u-v) \, R_{00''} (u-w) \, R_{0'0''} (v-w)  ~=~
  R_{0'0''} (v-w) \, R_{00''} (u-w) \, R_{00'} (u-v) \,  ,
\label {c3}
\eeq
where a matrix like $R_{ab}(u-v)$ acts nontrivially only on the $a$-th
and $b$-th vector spaces. The conserved quantities of HS spin chain
yield a realisation of the above mentioned 
monodromy matrix satisfying QYBE (\ref {c1}).

Now, for finding out algebraic relations among the conserved 
quantities (\ref {b11}), we consider an $R$ matrix of the form
\beq
R_{00'} (u-v)  ~=~   (u-v) \, Q_{00'} \, + \,  P_{00'} \, ,
\label {c4}
\eeq
where $Q_{00'}$ acts like $Q$ on the direct product of 
$0$-th and $0'$-th auxiliary spaces. Since this $Q$ matrix generates 
${\tilde P}_{ij}$ through the expression (\ref {a4}), it
is obvious that 
 there exists a one-to-one correspondence between 
the $R$ matrix (\ref {c4}) and the Hamiltonian (\ref {a7}) which 
contains ${\tilde P}_{ij}$. 
Moreover, by using the conditions (\ref {a5}) satisfied by $Q$ matrix, 
it is easy to check that the $R$ matrix (\ref {c4}) would be a solution of
  YBE (\ref {c3}). Consequently, 
by inserting the rational $R$ matrix (\ref {c4})
 to QYBE (\ref {c1}), one can construct
an associative, `extended Yangian'  algebra. At the limiting case
$Q_{00'} = \one $, (\ref {c4}) reduces to the $R$ matrix (\ref {c2}) 
 and leads to the standard $Y(gl_M)$ Yangian algebra.
However, in general, the $Q$ matrix appearing in (\ref {c4}) might also 
depend on a set of continuous as well as discrete deformation parameters. So
these parameters would naturally appear in the defining relations of
corresponding extended Yangian algebra. 
One may also make a short classification
of such extended Yangian algebras in the following way.
 Let us first assume that the 
$Q$ matrix, which is obtained by solving eqn.(\ref {a5}),
 admits a Taylor series expansion of the form (up to an over all sign factor)
\beq
Q_{00'} ~=~ \one ~+~ \sum_p \, h_p \, Q^p_{00'} ~+~ \sum_{p,q} \,
h_p h_q \, Q^{pq}_{00'} ~+~ \cdots \, ,
\label {c5}
\eeq
where $h_p$s are continuous deformation parameters and the leading
term is an identity operator.
Evidently the extended Yangian algebras, generated through this type of $Q$
 matrices, would reduce to  standard $Y(gl_M)$
algebra at the limit $h_p \rightarrow 0 $ for all $p$. 
So it is natural to call these algebras as 
multi-parameter dependent deformations of       
 $Y(gl_M)$ Yangian algebra. A few concrete examples of
 multi-parameter deformed Yangian algebras have already 
appeared in the literature [20-22] and it is recently found that
an integrable extension of Hubbard model respects such deformed Yangian 
symmetry [23].  However, it is also 
possible to find out some solutions of eqn.(\ref {a5}) 
which can not be expanded in the form (\ref {c5}) [17].
 So the Yangian algebras, generated through this type of
  $Q$ matrices and corresponding rational solutions
(\ref {c4}), would not reduce to  $Y(gl_M)$ algebra at
the limit $h_i \rightarrow 0$. Consequently,
these Yangians may be classified as  `nonstandard' variants
of $Y(gl_M)$ Yangian algebra.

At present our aim is to show that, the algebraic relations 
among the conserved quantities (\ref {b11}) 
would be given by an extended (i.e., a
multi-parameter deformed or `nonstandard' variant of) 
$Y(gl_M)$ Yangian.  To this end, we first consider the following  
 monodromy matrix whose operator valued elements act on the
Hilbert space of spin CS model (\ref {a1}) [5]:
\beq
T^0(u) ~=~ \Pi \left (  \, \one +  \, \sum_{i=1}^N
{P_{0i} \over u- D_i } \, \right ) \, ,
\label {c6}
\eeq
where  $D_i$s,  the so called Dunkl operators, are given by 
\beq
 D_i ~=~ \sum_{j \ne i } \, \theta_{ij} \,  K_{ij}  ~ ,
\label {c7}
\eeq
$ \theta_{ij} = { z_i \over z_i - z_j } $ and 
$K_{ij}$s, the coordinate
exchange operators,  are defined through the relations
\bea
&~~ \quad \quad  K_{ij} z_i ~=~ z_j K_{ij},~~K_{ij} { \d \over \d z_i } ~=~
{ \d \over \d z_j } K_{ij}, ~~K_{ij} z_l ~=~ z_l K_{ij}~,~ \nn &
\quad \quad  \quad   \quad  (3.8a)  \\
&~~ \quad \quad  K_{ij}^2 ~=~ 1, ~~~K_{ij}K_{jl} ~=~ K_{il}K_{ij} ~=~
K_{jl} K_{il} ~, ~~~[\, K_{ij}, K_{lm} \, ] ~=~ 0~,~ \nn  &
\quad  \quad \quad  \quad  (3.8b)
\eea
\addtocounter{equation}{1}
\noindent $ i, ~j,~l,~m ~$ being all different indices. Moreover,
the projection operator $\Pi$, appearing in (\ref {c6}),
 allows one to replace 
$K_{ij}$ by $P_{ij}$ (after taking $K_{ij}$ at the extreme right  
side of an expression).
It is easy to check that $D_i$s and $K_{ij}$s satisfy the 
commutation relations given by
\beq
K_{ij} D_i ~= ~ D_j K_{ij} \, , ~~\left [ \, K_{ij} , D_k \,
 \right ] ~=~ 0 \, , ~~\left [ \, D_i , D_j \, \right ] ~=~
 \left ( D_i -D_j \right ) K_{ij} \, ,
\label {c9}
\eeq
where $k\neq i,j $. By using these commutation relations, it can be 
 proved that the monodromy matrix (\ref {c6}) yields a solution of
QYBE (\ref {c1}) when the corresponding $R$ matrix is taken as
(\ref {c2}) [5].  

In analogy with (\ref {c6}), we now 
  propose another monodromy matrix as
\beq
T^0(u) ~=~ \Pi^* \left (  \, \Omega_0 +  \sum_{i=1}^N
{\P_{0i} \over u- D_i } \, \right ) \, ,
\label {c10}
\eeq
where 
\bea
~\Omega_0 ~ = ~ Q_{01} Q_{02} \cdots Q_{0N} \, , ~~
\P_{0i} ~ = ~ \left ( Q_{01} Q_{02} \cdots Q_{0,i-1} \right ) \, P_{0i} \,
\left ( Q_{0,i+1} Q_{0,i+2} \cdots Q_{0N} \right )   \, , \nn
~~~(3.11a,b)
\eea
\addtocounter{equation}{1}
and $\Pi^*$ denotes a new projection operator which 
allows one to replace $K_{ij}$ by ${\tilde P}_{ij}$ (after taking 
$K_{ij}$ at the extreme  right side of an expression):
$$
\Pi^* \left ( K_{ij} \right ) \, = \, {\tilde P}_{ij} \,= \,
\left( Q_{i,i+1}  Q_{i,i+2}  \cdots  Q_{ij} \right) \,  P_{ij} \, 
\left(  Q_{j-1,i} Q_{j-2,i}  \cdots   Q_{i+1,i}  \right) \, . 
$$
It is worth noting that the monodromy matrix (\ref {c10}), which 
reduces to (\ref {c6}) at the limiting case $ Q = \one $,
can be determined  uniquely from any given `anyon like'  
representation ${\tilde P}_{ij}$ (or, from the corresponding $Q$ matrix).
Furthermore, by  essentially following the approach of Ref.5 and using 
the relations (\ref {c9}) as well as (\ref {a5}), it can be
shown that the monodromy matrix (\ref {c10}) would satisfy QYBE (\ref {c1})
when the corresponding $R$ matrix is taken as (\ref {c4}).
Thus, we interestingly find that the monodromy matrix (\ref {c10})  yields 
a  realisation of extended $Y(gl_M)$ Yangian algebra. 
However, it should be noted that, 
the  operator valued elements of monodromy matrix 
(\ref {c10}) act on the Hilbert space of spin CS model, which contain 
both spin and dynamical degrees of freedom. So, for constructing operators
which would act only on the spin space,
 we consider a monodromy matrix like 
\beq
{\hat T}^0(u) ~=~ \l \, \omega_1 \omega_2 \cdots \omega_N \,
 \v \, T^0(u) \, \v \, \omega_1 \omega_2 \cdots \omega_N \, \r \, ,
\label {c12}
\eeq
where $T^0(u)$ is given by (\ref {c10}), and 
 $ \v  \omega_1 \omega_2 \cdots \omega_N  \r $ denotes a ket vector 
related to the `static limit' of spin CS model: 
$ z_k \v  \omega_1 \omega_2 \cdots \omega_N  \r = \omega^k 
 \v \omega_1 \omega_2 \cdots \omega_N  \r $. 
Since 
 $T^0(u)$  (\ref {c10}) does not depend on any momentum operator
like ${\d \over \d z_k }$,
it is clear that the monodromy matrix ${\hat T}^0(u)$ (\ref {c12}) 
would also give a realisation of extended $Y(gl_M)$ Yangian algebra 
generated by the $R$ matrix (\ref {c4}). Moreover, the elements of 
${\hat T}^0(u)$ would evidently act on the Hilbert space associated 
with HS like spin chain.

Next, we want to write down two  simple relations, 
which will be used shortly
 for establishing a connection between the conserved quantities (\ref {b11})
and our realisation (\ref {c12}) of extended Yangian algebra.
First of all, by applying the commutation relations (\ref {c9}), 
 it is easy to find  that
\beq
 \l \,  \omega_1 \omega_2 \cdots \omega_N \,  \v \,
\Pi^* \left ( D_i^n \right ) \,
 \v \, \omega_1 \omega_2 \cdots \omega_N  \, \r 
 ~=~ \sum_{j=1}^N \,
\left ( {\cal L}_{HS}^n  \right )_{ij} \, ,
\label {c13}
\eeq
where the elements of $ {\cal L}_{HS}$ are given by eqn.(\ref {b10}).
Secondly, by using the standard
relation: $ P_{0i} ~=~ \sum_{\alpha , \beta = 1}^M \,
X_0^{\alpha \beta } \otimes  X_i^{\beta \alpha } \,$
and the conditions (\ref {a5}) on $Q$ matrix,  the operator
$\P_{0i}$ (3.11b)  can be rewritten in a compact form like
\beq
\P_{0i} ~=~ \sum_{\alpha , \beta = 1}^M \,
X_0^{\alpha \beta } \otimes {\tilde X}_i^{\beta \alpha } \, .
\label {c14}
\eeq
Now, with the help of relations (\ref {c13}) and (\ref {c14}),
we can interestingly 
express the monodromy matrix (\ref {c12}) through its modes as
\beq
{\hat T}^0(u) ~=~ \Omega_0 \, + \,  \sum_{n=0}^\infty  \,
{1\over u^{n+1} } \, \sum_{\alpha ,\beta = 1}^M  \,
\left (\, X_0^{\alpha \beta } \otimes
 T_n^{\beta \alpha } \, \right )  ~ ,
\label {c15}
\eeq
 where $ T_n^{\alpha \beta }$s are the conserved quantities (\ref {b11})
which commute with Hamiltonian (\ref {a7}). 

It is curious to
notice that, apart from the conserved quantities (\ref {b11}) which 
 have been derived from the Lax equations, 
 the operator $\Omega_0$  also appears in the mode expansion of 
monodromy matrix (\ref {c15}).  But it seems that, by
 applying  Lax equations (\ref {b9}), it is not possible
 to determine whether this $\Omega_0$  would also commute 
with the spin chain Hamiltonian (\ref {a7}).
However, with the help of eqn.(\ref {a5}), one can directly show that
\beq
 \left [ \, \Omega_0  \, , \, {\tilde P}_{ij} \, \right ] ~=~ 
 \left [ \, \Omega^{\alpha  \beta }
  \, , \, {\tilde P}_{ij} \, \right ] ~=~ 0 \, ,  
\label {c16}
\eeq
where 
 $ \Omega^{ \alpha \beta }$s  are the operator valued elements of 
$\, \Omega_0$: $~ \Omega_0 \, 
= \sum_{\alpha , \beta }  X_0^{\alpha \beta }  \otimes 
  \Omega^{\beta \alpha } \, . $
Using the relation (\ref {c16}), we may now easily verify that  
$\Omega_0$  and $ \Omega^{ \alpha \beta }$
indeed commute with the Hamiltonian (\ref {a7}).  Thus  
we curiously observe that,  HS like spin chains possess 
some additional conserved quantities ($\Omega^{\alpha \beta }$s)
which are not apparently derivable from the Lax equations. However,  
it is obvious that  these extra conserved quantities would
become trivial at the limit $Q= \one$ and, therefore, one does not encounter
 such conserved quantities in the case of usual HS model.
From the above discussions it also turns out that, all modes
of the monodromy matrix ${\hat T}^0(u)$ (\ref {c15}) 
can be identified with various conserved 
quantities which commute with the Hamiltonian (\ref {a7}). 
Therefore, we may conclude that the HS like spin chain (\ref {a7})
respects an extended $Y(gl_M)$ Yangian symmetry. 
Furthermore, we can explicitly construct the algebraic relations among the
conserved quantities of such spin chain, 
by simply expressing this extended $Y(gl_M)$ Yangian algebra
through the modes of corresponding monodromy matrix. 

It is worth noting that, for proving the extended 
 $Y(gl_M)$ Yangian symmetry of Hamiltonian (\ref {a7}),
 we have not used so far any particular form of the related 
`anyon like' representation and only assumed
that such a representation can be obtained through the relation (\ref {a4}).
Thus, we are able to develop a rather general framework for finding out the 
conserved quantities and symmetry algebra associated  with all
 Hamiltonians  which can be expressed in the form (\ref {a7}).
In the following, however,  we like to discuss about some concrete examples 
of `anyon like' representations of permutation algebra, the related 
HS like spin chains  and corresponding symmetry algebras. To this end,
we first observe that their exists a class of solutions of 
eqn.(\ref {a5}) as
\beq
Q_{ik}~=~ \sum_{\sigma =1}^M \, e^{i\phi_{\sigma \sigma }}
 \, X_i^{ \sigma \sigma }\otimes X_{k}^{ \sigma \sigma }~+~
\sum _{ \sigma \neq \gamma } \, e^{i\phi_{\gamma \sigma }}
 \, X_i^{ \sigma \sigma }\otimes X_{k}^{ \gamma \gamma } \, ,
\label {c17}
\eeq
where $\phi_{\sigma \sigma }$s are some discrete parameters
each of which can be freely chosen as $0$ or $\pi$, and 
$~\phi_{ \gamma  \sigma }$s are 
 continuous deformation parameters which satisfy 
the antisymmetry property:
$ \phi_{ \gamma  \sigma }  =  - \phi_{ \sigma  \gamma } $.
So, for any possible choice of the above mentioned $M$ number of 
discrete parameters and ${ M(M-1) \over 2 }$ number of
independent continuous parameters,  
(\ref {c17}) would give us a distinct solution of (\ref {a5}). 
By substituting the $Q$ matrix solution (\ref {c17}) to (\ref {a4}), 
one can easily obtain a class of `anyon like' representations 
which will also depend on these discrete as well as continuous 
deformation parameters and  act on the spin space as 
\bea
&&{\tilde P}_{ij} \,
 \v \alpha_1 \alpha_2 \cdots \alpha_i \cdots \alpha_j
\cdots \alpha_N \r ~~=~~~~~~~~~~~~~~~~~~~~~~~~~~~~~ \nn \\ &&
 \quad \quad \quad
 \exp   \left \{  \,  i \, \phi_{\alpha_i \alpha_j} ~+~ i \,
\sum_{\tau =1}^M  \, n_\tau \,   \left (   \phi_{\tau \alpha_j} -
 \phi_{\tau \alpha_i}    \right )  \,  \right \}   \,
     \v \alpha_1
\alpha_2 \cdots \alpha_j \cdots \alpha_i \cdots \alpha_N \r   , ~~
\label {c18}
\eea
where $n_\tau $ denotes the number of times of occurring
 the  particular spin orientation $ \tau $
in the configuration $ \v \alpha_1  \cdots \alpha_i \cdots
\alpha_p \cdots \alpha_j \cdots  \alpha_N \r $,
when the index $p$ in $\alpha_p $ is varied from $i+1$ to $j-1$.
Evidently, for the trivial choice of deformation parameters as:
$ \phi_{ \gamma  \gamma } ~=~ \phi_{ \gamma  \sigma }~=~0$  
for all $\sigma ,\,  \gamma $,
 ${\tilde P}_{ij}$ (\ref {c18})  reduces to the standard 
representation of permutation algebra, which does not pick up any 
phase factor while interchanging the spins of two lattice points.
However, for any nontrivial choice of these deformation parameters, 
 the above ${\tilde P}_{ij}$  picks up a phase factor
which  would depend on the spin configuration
of $(j-i+1)$ number of lattice points.
Consequently, by substituting such ${\tilde P}_{ij}$ to (\ref {a7}),
we can generate concrete examples of HS like spin chain.
The Lax pairs and conserved quantities, associated with these spin chains,
may also be obtained 
by  inserting ${\tilde P}_{ij}$ (\ref {c18}) and the related  
$ {\tilde X}_i^{ \alpha \beta }$ (\ref {b3}) to the expressions
(\ref {b10}) and (\ref {b11}).
 Moreover,  the  form of $Q$ matrix (\ref {c17}) suggests 
that $\Omega_0$ (3.11a) can now be written in a diagonal form:
 $ \Omega_0 \, 
~=~  \, \sum_{\alpha  } \, X_0^{\alpha \alpha } \, \otimes 
 \, \Omega^{\alpha \alpha } \, $. So these 
 $ \Omega^{\alpha \alpha }$s would give us 
 $M$ number of additional conserved quantities which are not related 
to the Lax equations. 

Now, for finding out algebraic relations 
 among the above mentioned conserved quantities, 
we substitute the $Q$
matrix (\ref {c17}) to (\ref {c4}) and  obtain
a class of rational $R$ matrices which also 
depend on the parameters $\phi_{\sigma \sigma },~ \phi_{\sigma \gamma }$ : 
\beq
R_{00'} (u-v)  ~=~   (u-v) \,
\sum _{ \sigma , \gamma = 1}^M  \, e^{i\phi_{\gamma \sigma }}
 \, X_0^{ \sigma \sigma }\otimes X_{0'}^{ \gamma \gamma } \, + \,
\sum_{ \sigma , \gamma =1 }^M   \, 
  X_0^{ \sigma \gamma }\otimes X_{0'}^{ \gamma \sigma } \, .
\label {c19}
\eeq
It is clear that, by inserting these $R$ matrices
to QYBE (\ref {c1}), one can generate a class of extended 
$Y(gl_M)$ Yangian algebras. However it should be noted that,
 only for the special choice of discrete parameters as
$\phi_{\sigma \sigma } = 0 $ (or, $\phi_{\sigma \sigma } = \pi $)
 for all $\sigma $,
the $Q$ matrix (\ref {c17}) admits an expansion in the form
(\ref {c5}). Therefore, only for these two choices of discrete
parameters, the corresponding $R$ matrices (\ref {c19}) would generate
 multi-parameter dependent deformations of $Y(gl_M)$ Yangian
algebra. For any other choice of discrete parameters
$\epsilon_{\sigma } $, the $R$ matrix (\ref {c19}) 
will evidently lead to a `nonstandard' variant of $Y(gl_M)$ Yangian algebra. 
The monodromy matrix (\ref {c15}), associated with the representation
(\ref {c18}), would give us concrete realisation of such
multi-parameter deformed or `nonstandard' variant 
of $Y(gl_M)$ Yangian algebra.  By inserting 
this monodromy matrix (\ref {c15}) as well as 
 $R$ matrix (\ref {c19}) to QYBE (\ref {c1}), and equating from its both 
sides the coefficients of same  powers in  $u,~v$, we easily 
obtain the following set of relations:
\bea
     &\left[\, \Omega^{\alpha \alpha } \, , \, \Omega^{\beta \beta } \, 
\right ] ~=~0~,~~~~ \Omega^{\alpha \alpha } \, T_n^{ \beta \gamma } ~=~ 
{\rho_{ \alpha \gamma } \over \rho_{\alpha \beta }} \,
T_n^{ \beta \gamma } \Omega^{\alpha \alpha } \, , ~~
\nn &(3.20a,b) ~~~~~~~~
\\
 &\rho_{ \alpha \gamma } \, T^{\alpha \beta }_0 \, T^{ \gamma \delta }_n 
~-~\rho_{\beta \delta } \, T^{\gamma \delta }_n  \, T^{\alpha \beta }_0 
~~=~~
\delta_{ \alpha \delta } \, T^{\gamma \beta }_n \, 
\Omega^{\alpha \alpha  }~-~\delta_{ \gamma \beta }
~\Omega^{ \gamma \gamma } \, T^{\alpha \delta }_n \,,~ \nn  &~~(3.20c)~~~~~~~~
 \\
 &\left [ ~\rho_{ \alpha \gamma } \, T^{ \alpha \beta }_{n+1} \, 
 T^{\gamma \delta }_m  \, - \, \rho_{\beta \delta } \,
T^{ \gamma \delta }_m \, T^{ \alpha \beta }_{n+1}~ \right ] ~-~
 \left [ ~\rho_{ \alpha \gamma } \, T^{ \alpha \beta }_{n} \, 
 T^{\gamma \delta }_{m+1}  \, - \, \rho_{\beta \delta } \,
T^{ \gamma \delta }_{m+1} \, T^{ \alpha \beta }_n~ \right ]  \nn
&\\ &\hskip 8 cm ~=~   ~\left (~ T^{ \gamma \beta }_m \,
T^{ \alpha \delta }_n  ~-~  T^{ \gamma \beta }_n \, T^{ \alpha \delta }_m~
\right ) \, , \nn &~~(3.20d) ~~~~~~~~
\eea
\addtocounter{equation}{1}
where $\rho_{ \alpha \gamma } = e^{ i\phi_{ \alpha \gamma }} $. Thus 
we are able to derive here the algebraic relations among the conserved 
quantities of spin chain (\ref {a7}) containing `anyon like' 
representation (\ref {c18}). It may be noted that, at the special case
 $\rho_{ \alpha \gamma } = 1 $ and $\Omega^{\alpha \alpha } = \one $ 
for all $\alpha , ~ \gamma $, the algebra (3.20)
reduces to standard $Y(gl_M)$ Yangian and reproduces the 
commutation relations between the conserved quantities of HS spin chain.

Finally we like to mention that the quantum $R$ matrix 
(\ref {c4}), which played a crucial role in finding out
the symmetry of  HS like spin chain (\ref {a7}), also
appeared previously in the context of integrable models 
with short ranged interaction.  As it is  well known, 
one can construct a generalisation of symmetric six-vertex model by
introducing horizontal and vertical electric fields  which interact with
the related dipoles [24]. Higher dimensional extensions 
of such asymmetric six vertex
model, containing $M$-state ($M>2$) bonds at each vertex point,
 are also found [25]. Furthermore, one can obtain an 
integrable generalisation of Heisenberg spin chain by introducing the
Dzyaloshinsky-Moriya (DM) interaction [26]. However, it may be noted
that, the rational limit of all quantum $R$-matrices, 
which appear in the algebraic Bethe ansatz of 
Heisenberg spin chain with DM interaction, the asymmetric
six vertex model and its higher dimensional extensions,
can always  be written in the form  (\ref {c4}). So,
 there might exist an intriguing connection between the algebraic 
structures of the above mentioned integrable models with short ranged
 interactions and the HS like spin chains (\ref {a7}) considered by us.

\vspace{1cm}

\noindent \section { Concluding Remarks }

\medskip
Here we have constructed some novel variants of 
Haldane-Shastry (HS)  spin chain which  would respect 
extended (i.e., multi-parameter dependent or `nonstandard')
$Y(gl_M)$ Yangian symmetries.
An interesting feature of these spin chains is that they
contain nonlocal interactions, which can  be expressed
through the `anyon like' representations of permutation algebra.
We have also established the integrability of such spin chains,  
by  finding out the  associated Lax pairs and conserved quantities.
However, it turned out that these models also possess a few additional 
conserved quantities which are not derivable from the Lax
equations.  These additional conserved quantities  are found to 
play an important role in generating the monodromy matrix and 
 symmetry algebra of the above mentioned HS like spin chains.

The existence of extended $Y(gl_M)$ Yangian symmetry in integrable variants
of HS spin chain might lead to further developments in several directions.
For example, it should be interesting to investigate whether
 these spin chains, containing `anyon like' representations of 
permutation algebra, can  also be solved exactly in analogy with
their standard counterpart. However, it is found earlier that,
 the spectra of some spin CS models (\ref {a6}), with
  `nonstandard' Yangian symmetries,  differ significantly
 from that of the usual spin CS model (\ref {a1}) [16-17]. 
So it might be particularly interesting to search for the spectra of
HS like spin chains (\ref {a7}), which exhibit 
`nonstandard' Yangian symmetries.  Moreover, as it is well known,
the degeneracy of wave functions for HS model
can be explained quite nicely through the representations of
$Y(gl_M)$ algebra. Therefore, it is natural to expect that the 
representations of extended $Y(gl_M)$ algebras would play a similar role
in analysing the spectra of related HS like spin 
chains.  Furthermore, it might be fruitful  to apply 
asymptotic Bethe ansatz method for investigating various thermodynamic 
properties of these spin chains and explore their connection with 
Haldane's generalised exclusion statistics. Finally, we hope that
it would be possible to find out many other new type of quantum
integrable spin chains and dynamical models,
which would exhibit the extended $Y(gl_M)$ Yangian symmetries.

\medskip
\noindent {\bf Acknowledgments }

 I am grateful to Prof. Miki Wadati 
 for many illuminating discussions and critical reading of the 
manuscript. I like to thank Drs. K. Hikami, H. Ujino and M. Shiroishi
for many fruitful discussions. This work is supported by a grant
(JSPS-P97047) of the Japan Society for the Promotion of Science.

\newpage
\leftline {\large \bf References }
\medskip
\begin{enumerate}

\item  F.D.M. Haldane, Phys. Rev. Lett. 60 (1988) 635; \hfil \break 
       B.S. Shastry, Phys. Rev. Lett. 60 (1988) 639.

\item  F.D.M. Haldane, Z.N.C. Ha, J.C. Talstra, D. Bernerd and V. Pasquier,
       Phys. Rev. Lett. 69 (1992) 2021.

\item  A.P. Polychronakos, Phys. Rev. Lett. 69 (1992) 703.

\item  B. Sutherland and B.S. Shastry, Phys. Rev. Lett. 71 (1993) 5.

\item  D. Bernard, M. Gaudin, F.D.M. Haldane and V. Pasquier,
       J. Phys. A 26 (1993) 5219.

\item  Z.N.C. Ha and F.D.M. Haldane, Bull. Am. Phys. Soc. 37 (1992) 646;
       \hfil \break 
       N. Kawakami, Phys. Rev. B46 (1992) 1005.

\item  K. Hikami and M. Wadati, J. Phys. Soc. Jpn. 62 (1993) 4203;
       Phys. Rev. Lett. 73 (1994) 1191.

\item  H. Ujino and M. Wadati, J. Phys. Soc. Jpn. 63 (1994) 3585;
       J. Phys. Soc. Jpn. 64 (1995) 39.

\item  K. Hikami, Nucl. Phys. B 441 [FS] (1995) 530.

\item  F.D.M. Haldane,{\it in} Proc. 16th Taniguchi Symp., Kashikijima,
       Japan, (1993) eds. A. Okiji and N. Kawakami (Springer, Berlin,
       1994).

\item  A. Cappelli, C.A. Trugenberger, and G.R. Zemba, Phys. Rev. Lett.
       72 (1994) 1902.

\item  M. Stone and M. Fisher, Int. J. Mod. Phys. B 8 (1994) 2539.

\item  F. Lesage, V. Pasquier and D. Serban, Nucl. Phys. B 435 [FS] (1995)
       585.

\item  H. Awata, Y. Matsuo, S. Odake and J. Shiraishi, Nucl. Phys. B 449
       (1995) 347.

\item  J. Avan, A. Jevicki, Nucl. Phys. B 469 (1996) 287.

\item  B. Basu-Mallick, Nucl. Phys. B 482 [FS] (1996) 713.

\item  B. Basu-Mallick and A. Kundu, Nucl. Phys. B 509 [FS] (1998) 705.

\item  V.G. Drinfeld, {\it Quantum Groups }, in ICM proc. (Berkeley, 1987)
       p. 798.

\item  V. Chari and A. Pressley, {\it A Guide to Quantum Groups} (Cambridge
       Univ. Press, Cambridge, 1994).

\item  B. Basu-Mallick and P. Ramadevi, Phys. Lett. A 211 (1996) 339.

\item  B. Basu-Mallick, P. Ramadevi and R. Jagannathan, 
       Int. Jour. Mod. Phys. A 12 (1997) 945.

\item  A. Stolin and P.P. Kulish, {\it New rational solution
       of Yang-Baxter equation and deformed Yangians }, TRITA-MAT-1996-JU-11,
       q-alg/9608011.

\item  A. Kundu, {\it Exactly integrable family of generalised Hubbard
       models with twisted Yangian symmetry}, cond-mat/9710033.

\item  C.P. Yang, Phys. Rev. Lett. 19 (1967) 586; \hfil \break
       B. Sutherland, C.N. Yang, C.P. Yang, Phys. Rev. Lett. 19 (1967)
       588.

\item  J.H.H. Perk and C.L. Schultz, {\it in} Yang-Baxter equation 
       in integrable systems, ed. M. Jimbo, Advanced Series in Math.
       Phys., Vol.10 (World Scientific, Singapore, 1990), p.326.

\item  F.C. Alcaraz and W.F. Wreszinski, J. Stat. Phys. 58 (1990) 45.

\end{enumerate}
\end{document}